\documentclass[a4paper]{article}

\usepackage[english]{babel}
\usepackage[utf8]{inputenc}
\usepackage{amsmath}
\usepackage{graphicx}
\usepackage[affil-it]{authblk}
\usepackage{apalike}
\usepackage[UKenglish]{isodate}
\usepackage{url}

\title{Analysis of family-wise error rates in statistical parametric mapping using random field theory}

\author{Guillaume Flandin and Karl J. Friston}
\affil{Wellcome Trust Centre for Neuroimaging\\ University College London, UK}

\date{\today}

\begin{document}
\maketitle

\begin{abstract}

This technical report revisits the analysis of family-wise error rates in statistical parametric mapping -- using random field theory -- reported in \cite{eklund15}. Contrary to the understandable spin that these sorts of analyses attract, a review of their results suggests that they endorse the use of parametric assumptions -- and random field theory -- in the analysis of functional neuroimaging data. We briefly rehearse the advantages parametric analyses offer over nonparametric alternatives and then unpack the implications of \cite{eklund15} for parametric procedures.

\end{abstract}

\section{Introduction}
\label{sec:introduction}

Random field theory has been at the heart of statistical parametric mapping in neuroimaging -- and its various implementations in academic software -- for over two decades. With technical advances in data acquisition, its validity has been revisited every few years to ensure it is fit for purpose \cite{worsley96,hayasaka03,hayasaka04,pantazis05,bennett09,nichols12,woo14}, particularly in relation to controlling family-wise error. The statistical validity of procedures based on random field theory is important because random field theory offers an efficient and reproducible alternative to nonparametric testing. The advantages of parametric approaches over nonparametric approaches include the following:

\begin{itemize}

\item Parametric approaches are more efficient than their nonparametric counterpart by the Neyman-Pearson lemma. This follows because the most efficient test is based upon the odds ratio inherent in parametric tests. This means that any nonparametric test can only be as efficient as a parametric test or less efficient.

\item Parametric approaches are reproducible. In other words, one obtains the same result when repeating the analysis, unlike the \emph{p}-values based upon samples of the null distribution used in nonparametric tests.

\item Parametric approaches eschew the problem of complying with the exchangeability criteria of nonparametric procedures. These criteria make it difficult to apply nonparametric tests to data that have serial correlations or when using hierarchical models.

\item Parametric approaches are computationally more efficient because they use distributional assumptions to eschew computationally intensive sampling from a null distribution.

\end{itemize}

These advantages rest upon distributional assumptions that, if violated, render parametric tests inexact. In these instances, one could consider using nonparametric tests. It is therefore important to ensure that parametric tests and random field theory are robust to any violations. The analyses reported by \cite{eklund15} speak to this issue. So what conclusions can be drawn from these analyses?

\section{A review of the Eklund et al simulation results}
\label{sec:review}

\cite{eklund15} assess the family-wise error rate using parametric and nonparametric tests and a variety of regressors to analyse (publicly available) resting state fMRI data from two sites. They manipulate a number of factors including: (i) inference based upon peak height versus spatial extent; (ii) spatial extent inference based upon high versus low cluster forming thresholds; (iii) under different levels of spatial smoothing for (iv) block versus event-related regressors, using (v) one- and two-sample \emph{t}-tests.

In brief, they show that parametric inference based upon peak height is well-behaved and provides acceptable family-wise error control. In contrast, parametric inference based upon spatial extent is not valid when, and only when, a low cluster forming threshold is employed. This failure is well known and is consistent with random field theory: the null distribution for spatial extent is based upon the Nosko conjecture that provides a distributional form for the spatial extent of a cluster \cite{friston94}. The parameter of this distributional form is fixed using approximations to the expected number of maxima and the total volume above a threshold (see \cite{flandin15} for a brief review). Crucially, both the distributional form for the spatial extent and the expected number of maxima (the Euler characteristic) are approximations that are only true in the limit of high thresholds (see Figure 1 in \cite{friston94}). This means that tests based upon spatial extent become inexact at low thresholds -- as verified numerically by \cite{eklund15}.

The effects of smoothing reported in \cite{eklund15} are consistent with random field theory, which assumes a good lattice approximation to a continuous random field. This assumption means that the data have to be smoother than the size of voxels. In other words, increasing the smoothness will lead to more exact inference. Again, this is verified numerically by \cite{eklund15}.

The effect of one versus two-sample \emph{t}-tests is slightly more difficult to interpret. This is because the authors used the same regressor for all subjects. Arguably, this was a mistake because any systematic fluctuation in resting state timeseries -- that correlates with the regressor -- will lead to significant one-sample \emph{t}-tests against the null hypothesis of zero (e.g., magnetic equilibration effects). This effect is particularly marked for a regressor (called E1) that represents a fast and inefficiently estimated event-related response every few seconds. Crucially, the nonparametric false positive rates are beyond the 95\% confidence intervals. This means that this effect is actually expressed in the data over subjects and therefore fails as a model of the null behaviour.

\begin{figure}
\centering
\includegraphics[width=1\textwidth]{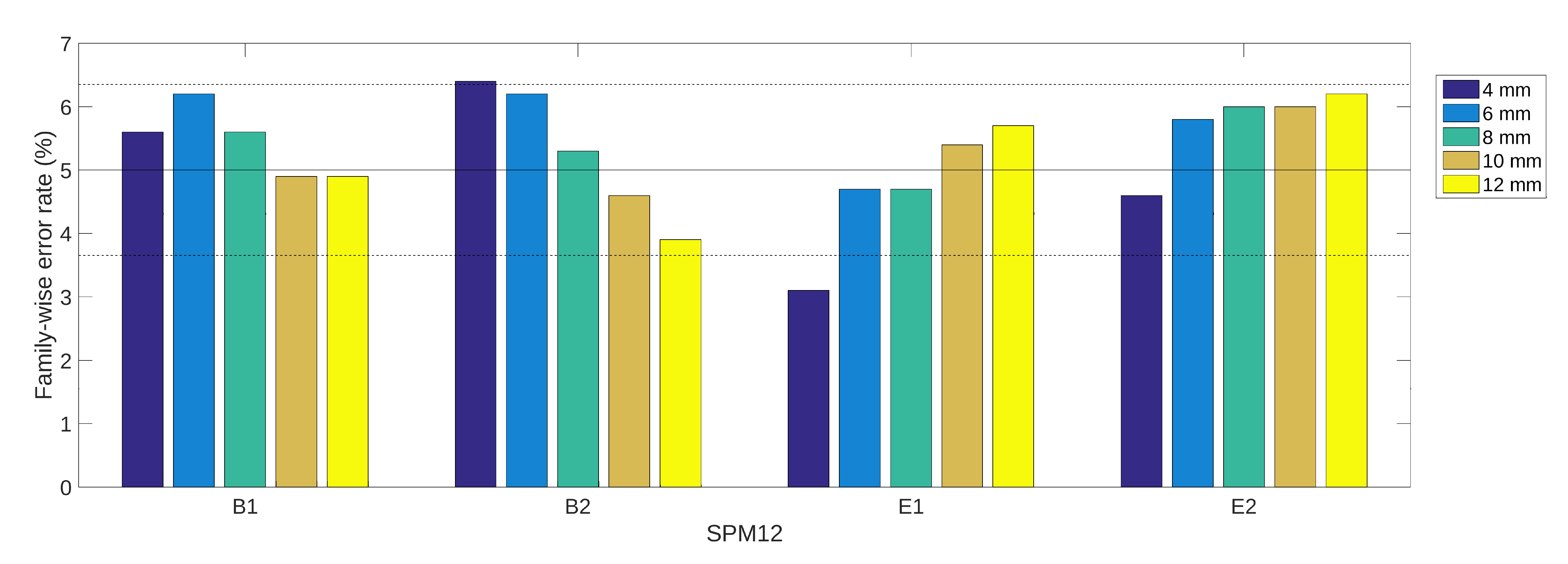}
\caption{\label{fig:fig1} Cluster-level inference results for a two-sample \emph{t}-test (two groups of ten random subjects, repeated a thousand times) with the Beijing dataset using a cluster-forming threshold of $p<0.001$ and the SPM12 software (r6685). Five levels of spatial smoothing were evaluated (4, 6, 8, 10 and 12 mm isotropic Gaussian kernels) with four different regressors (see \cite{eklund15} for details).}
\end{figure}

This failure is finessed when comparing parameter estimates between two groups using a two-sample \emph{t}-test. In this instance, inferences based upon spatial extent fall to acceptable family-wise error rates. We confirmed this by reproducing the analysis (using the same data and regressors) reported in \cite{eklund15} (see Figure \ref{fig:fig1}). These analyses use the close to original (3 mm) voxels sizes -- as opposed to the upsampled (2 mm voxel) data as analysed in \cite{eklund15}.

\section{Conclusion}
\label{sec:conclusion}

The results provided by \cite{eklund15} endorse the use of random field theory and reveal behaviour that is consistent with the underlying distributional assumptions. Having said this, there was a problem with the construction of the null distributions presented in \cite{eklund15}. This is because they used the same regressors for every subject. However, this problem is finessed by the use of two-sample \emph{t}-tests. The results of these analyses (two-sample \emph{t}-test at high thresholds) show that the random field theory provides valid inference based on spatial extent, provided its distributional assumptions are not violated (through the use of low cluster forming thresholds or smoothing). This conclusion is important because the issues addressed in \cite{eklund15} underwrite good practice in imaging neuroscience.

\section{Acknowledgements}
\label{sec:acknowledgements}

We would like to thank our colleagues Chris Mathys and Will Penny for their help in the preparation of this report. We are also grateful to the 1000 Functional Connectomes Project\footnote{\url{http://fcon_1000.projects.nitrc.org/}}, the Neuroimaging Informatics Tools and Resources Clearinghouse (NITRC) and the corresponding researchers for making publicly available the resting state datasets used in our analyses.

\bibliography{biblio}
\bibliographystyle{apalike}  

\end{document}